\documentclass[preprint,12pt]{elsarticle}

\usepackage{amssymb}
\usepackage[nointegrals]{wasysym}

 \usepackage{amsthm}
 \usepackage{amsmath}
\usepackage{xcolor}
\definecolor{silver}{rgb}{0.75, 0.75, 0.75}

 \usepackage{lineno}
 \theoremstyle{plain}
 \newdefinition{definition}{Definition}
\newtheorem{theorem}{Theorem}
 
  \newtheorem{corollary}[theorem]{Corollary}
 \newdefinition{remark}{Remark}
 \newdefinition{property}{Property}

\journal{arXiv}

\begin{document}

\begin{frontmatter}



\title{Optimal discrimination design for copula models}

\author[IST]{E. PERRONE }
\ead{elisa.perrone@ist.ac.at}
\author[JKU]{A. RAPPOLD}
\ead{andreas.rappold@jku.at}
\author[JKU]{W.G. M\"ULLER}
\ead{werner.mueller@jku.at}

\address[IST]{Institute of Science and Technology, 3400 Klosterneuburg, Austria}
\address[JKU]{Department of Applied Statistics, Johannes Kepler University of Linz,\\ 4040 Linz, Austria}

\begin{abstract}
Optimum experimental design theory has recently been extended for parameter estimation in copula models. However, the choice of the correct dependence structure still requires wider analyses. In this work the issue of copula selection is treated by using discrimination design techniques. The new proposed approach consists in the use of $D_s$-optimality following an extension of corresponding equivalence theory. We also present some examples and highlight the strength of such a criterion as a way to discriminate between various classes of dependences.
\end{abstract}


\end{frontmatter}


\section{Introduction}

One of the most important tasks in copula modeling is to decide which specific copula to employ. For that purpose a rather general approach is to use omnibus goodness-of-fit tests that require minimum assumptions, for recent reviews see, e.g., \cite{Berg_2009}, \cite{Genest_et_al_2009}, or \cite{Fermanian_2013}. Other more specific avenues consist in applying graphical tools (\cite{michiels+d_13}) or information based criteria (\cite{gronneberg+h_14}). In fully parametric models, as considered in this paper, the latter can be formulated in terms of functions of the Fisher information matrices, which will allow us to generate optimal designs for copula model discrimination.  

Design optimization is generally largely employed in many applied fields as a convenient tool to improve drawing informative experiments. 
Recently, in \cite{Perrone_16} the theory of $D$-optimality has been extended to a wider class of models for the usage of copulas. Although the employment of such functions allows for a substantial flexibility in modeling, it also leads to the natural question of their (proper) choice. As stated,
developments of powerful goodness-of-fit tests and strategies to avoid the wrong choice of the dependence constitute a considerable part of the literature on copulas. The issue of model choice or discrimination is in principle also a well known part of (optimum) experimental design theory and several criteria (e.g., $D_s$-optimality, $T$-optimality, $KL$-optimality) have been proposed (see \cite{dette+t_09,lopez-fidalgo_2007,studden_80}, and \cite{deldossi_16} for a special application to copula models).  

In this work we first extend the general theory of $D_A$-optimality to copula models. Then, we present the usage of the $D_s$-criterion to discriminate between various classes of dependences and possible scenarios. Finally, we show through some examples possible real applications.

\section{Theoretical framework}
In this section we provide the extension for the $D_A$-criterion of a Kiefer-Wolfowitz type equivalence theorem, assuming the dependence described by a copula model. We then illustrate the basic idea of the new approach through a motivating example already analyzed in \cite{Perrone_16}.
\subsection{$D$-, $D_A$-, and $D_s$-optimality}
Let us consider a vector $\mathbf{x}^T = (x_1, \ldots, x_r) \in \mathcal{X}$ of control variables, where
$\mathcal{X} \subset \mathbb{R}^r$ is a compact set. The results of the observations and of the expectations in a regression experiment are the vectors
$$\mathbf{y}(\mathbf{x}) = (y_1(\mathbf{x}), y_2(\mathbf{x})),$$
\[\mathbf{E}[\mathbf{Y}(\mathbf{x})] = \mathbf{E}[(Y_1,Y_2)] = \boldsymbol{\eta}(\mathbf{x},\boldsymbol{\beta}) = (\eta_1(\mathbf{x},\boldsymbol{\beta}),\eta_2(\mathbf{x},\boldsymbol{\beta})),\]
where $\boldsymbol{\beta}=(\beta_1, \ldots,\beta_k)$ is a certain unknown parameter vector to be estimated and $\eta_i \; (i = 1,2)$ are known functions.

Let us call $F_{Y_i}(y_i(\mathbf{x}, \boldsymbol{\beta}))$ the margins of each $Y_i$ for all $i\in\{1,2\}$ and $f_{\mathbf{Y}}(\mathbf{y}(\mathbf{x}, \boldsymbol{\beta}), \boldsymbol{\alpha})$ the joint probability density function of the random vector $\mathbf{Y}$, where $\boldsymbol{\alpha}=({\alpha}_1,\ldots, {\alpha}_l)$ is the unknown copula parameter vector.

According to Sklar's theorem (\cite{nelsen_06}), let us assume that the dependence between $Y_1$ and $Y_2$ is modeled by a copula function $$C_{\boldsymbol{\alpha}}(F_{Y_1}(y_1(\mathbf{x}, \boldsymbol{\beta})), F_{Y_2}(y_2(\mathbf{x}, \boldsymbol{\beta}))).$$ 

The Fisher Information Matrix $m(\mathbf{x}, \boldsymbol{\beta}, \boldsymbol{\alpha})$ for a single observation is a $(k +l) \times (k +l)$ matrix whose elements are 
\begin{equation}\label{Eq:FIM}
 \mathbf{E} \left(  - \dfrac{\partial^2}{\partial \gamma_i \partial \gamma_j} \log \Big[ \dfrac{\partial^2}{\partial y_1 \partial y_2} C_{\alpha}(F_{Y_1}(y_1(\mathbf{x}, \boldsymbol{\beta})), F_{Y_2}(y_2(\mathbf{x}, \boldsymbol{\beta})))\Big] \right)
 \end{equation}
where $\boldsymbol{\gamma}=\{{\gamma}_1,\ldots,{\gamma}_{k+l}\}=\{{\beta}_1,\ldots,{\beta}_k,{\alpha}_1,, \ldots, {\alpha}_l\}$.

The aim of design theory is to quantify the amount of information on both sets of parameters $\boldsymbol{\alpha}$ and $\boldsymbol{\beta}$, respectively, from the regression experiment embodied in the Fisher Information Matrix.

For a concrete experiment with $N$ independent observations at $n \le N$ support points $\mathbf{x_1},\ldots,\mathbf{x_n}$, the corresponding information matrix $M(\xi, \boldsymbol{\gamma})$ then is
\[M(\xi, \boldsymbol{\gamma}) = N^{-1} \sum\limits_{i=1}^n w_i \; m(\mathbf{x_i},\boldsymbol{\gamma}),\] 
where $w_i$ and $\xi$ are such that:
\[\sum\limits_{i=1}^n w_i = 1, \quad 
\xi = \left \{
\begin{array}{cccc}
\mathbf{x_1} &  \ldots & \mathbf{x_n} \\
w_1 &  \ldots & w_n
\end{array}
\right \}.
\]
The approximate design theory is concerned with finding $\xi^*(\boldsymbol{\gamma})$ such that it maximizes some scalar function $\phi(M(\xi,\boldsymbol{\gamma}))$, i.e., the so-called design criterion.

In \cite{Perrone_16}, we have developed the theory for the well known criterion of $D$-optimality, i.e., the criterion $\phi (M(\xi,\boldsymbol{\gamma})) = \log \det M(\xi,\boldsymbol{\gamma}) $, if $M(\xi,\boldsymbol{\gamma})$ is non-singular. 
In this work, we consider the case when the primary interest is in certain meaningful parameter contrasts. Such contrasts are element of the vector $A^T\boldsymbol{\gamma}$, where $A^T$ is an $s \times (k+l)$ matrix of rank $s < (k+l)$. 
If $M(\xi, \boldsymbol{\gamma})$ is non-singular, then the variance matrix of the least-square estimator of $A^T\boldsymbol{\gamma}$ is proportional to $A^T \{ M(\xi, \boldsymbol{\gamma}) \}^{-1} A$ and then a natural criterion, generalization of the $D$-optimality for this context, would be of maximizing $\log \det[A^T \{ M(\xi, \boldsymbol{\gamma}) \}^{-1} A]^{-1}$. 
This criterion is called \emph{$D_A$-optimality} (\cite{silvey_80}).

The following Theorem shows a generalization for the $D_A$-optimality of the Kiefer-Wolfowitz type equivalence theorem already proved in \cite{Perrone_16} for $D$-optimality. We have omitted the proof as it is, albeit a little more elaborate, fully analogous. 
\begin{theorem}
\label{Th:1}
For a localized parameter vector $(\tilde{\boldsymbol{\gamma}})$, the following properties are equivalent:
\begin{enumerate}
 \item $\xi^*$ is $D_A$-optimal;
\item  for every $\mathbf{x} \in \mathcal{X}$, the next inequality holds:
$$\textnormal{ tr }[ M(\xi^*, \tilde{\boldsymbol{\gamma}})^{-1} A (A^T M(\xi^*, \tilde{\boldsymbol{\gamma}})^{-1} A)^{-1} A^T M(\xi^*, \tilde{\boldsymbol{\gamma}})^{-1} m(\mathbf{x}, \tilde{\boldsymbol{\gamma}})]\leq s;$$  
\item over all $\xi \in \Xi$, the design $\xi^*$ minimizes the function
$$\max\limits_{x \in \mathcal{X}}\textnormal{ tr }[M(\xi^*, \tilde{\boldsymbol{\gamma}})^{-1} A (A^T M(\xi^*, \tilde{\boldsymbol{\gamma}})^{-1} A)^{-1} A^T M(\xi^*, \tilde{\boldsymbol{\gamma}})^{-1} m(\mathbf{x}, \tilde{\boldsymbol{\gamma}})].$$
\end{enumerate}
\end{theorem}
Although we here extend the theory to the general case of $D_A$-optimality, in the following our interest is in the first $s < (k+l)$ parameters, only. In such a case, $M(\xi, \boldsymbol{\gamma})$ can be written as:
$$M(\xi, \boldsymbol{\gamma}) = \left (
\begin{array}{cc}
M_{11} &  M_{12} \\
M_{12}^T &  M_{22}
\end{array}
\right ),$$
where $M_{11}$ is the $(s \times s)$ minor related to the estimated parameters. Therefore, the simplified criterion is to maximize the function $\phi_s (M(\xi, \boldsymbol{\gamma})) = \log \det (M_{11} - M_{12}M_{22}^{-1}M_{12}^T)$, which is called \emph{$D_s$-optimality}. We now have
\begin{corollary}
\label{Co:2}
 $D_s$-optimality follows as a particular case of Theorem \ref{Th:1} by the choice $A^T = (I_s \; 0)$. 
\end{corollary}
Given the characterization of Corollary \ref{Co:2}, two designs $\xi$ and $\xi^*$ can be compared by means of a ratio called \emph{$D_s$-Efficiency} defined as follows:
\[
\left(\dfrac{\det[M_{11}(\xi, \tilde{\boldsymbol{\gamma}}) - M_{12}(\xi, \tilde{\boldsymbol{\gamma}})M_{22}^{-1}(\xi, \tilde{\boldsymbol{\gamma}})M_{12}^T(\xi, \tilde{\boldsymbol{\gamma}})]}{\det[
M_{11}(\xi^*, \tilde{\boldsymbol{\gamma}}) - M_{12}(\xi^*, \tilde{\boldsymbol{\gamma}})M_{22}^{-1}(\xi^*, \tilde{\boldsymbol{\gamma}})M_{12}^T(\xi^*,\tilde{\boldsymbol{\gamma}})]}\right)^{1/s}.
\] 
In the next section we will describe the usage of $D_s$-optimality in the sense of discrimination through a simple example originally reported in \cite{fedorov_71}. 

\subsection{A motivating example}

Let us assume that for each design point $x \in [0,1]$, we observe an independent pair of random variables $Y_1$ and $Y_2$, such that
\[ E[Y_1(x)] = \beta_1 + \beta_2 x + \beta_3 x^2 ,\]
\[E[Y_2(x)] = \beta_4 x + \beta_5 x^3 + \beta_6 x^4.\]
The model is then linear in the parameter vector $\boldsymbol{\beta}$  and has dependence described by the product copula with Gaussian margins.

\begin{figure}[!ht]
\begin{minipage}{0.45\textwidth}
\centering
\begin{tabular}{r|r}
  $x$ & $w$ \\ 
  \hline
  0.0000 & 0.1502 \\ 
  0.3414 & 0.0854 \\ 
  0.7901 & 0.3419 \\ 
  1.0000 & 0.4226 \\ 
\end{tabular}
\end{minipage}
\begin{minipage}{0.45\textwidth}
\centering
\includegraphics[scale=0.38]{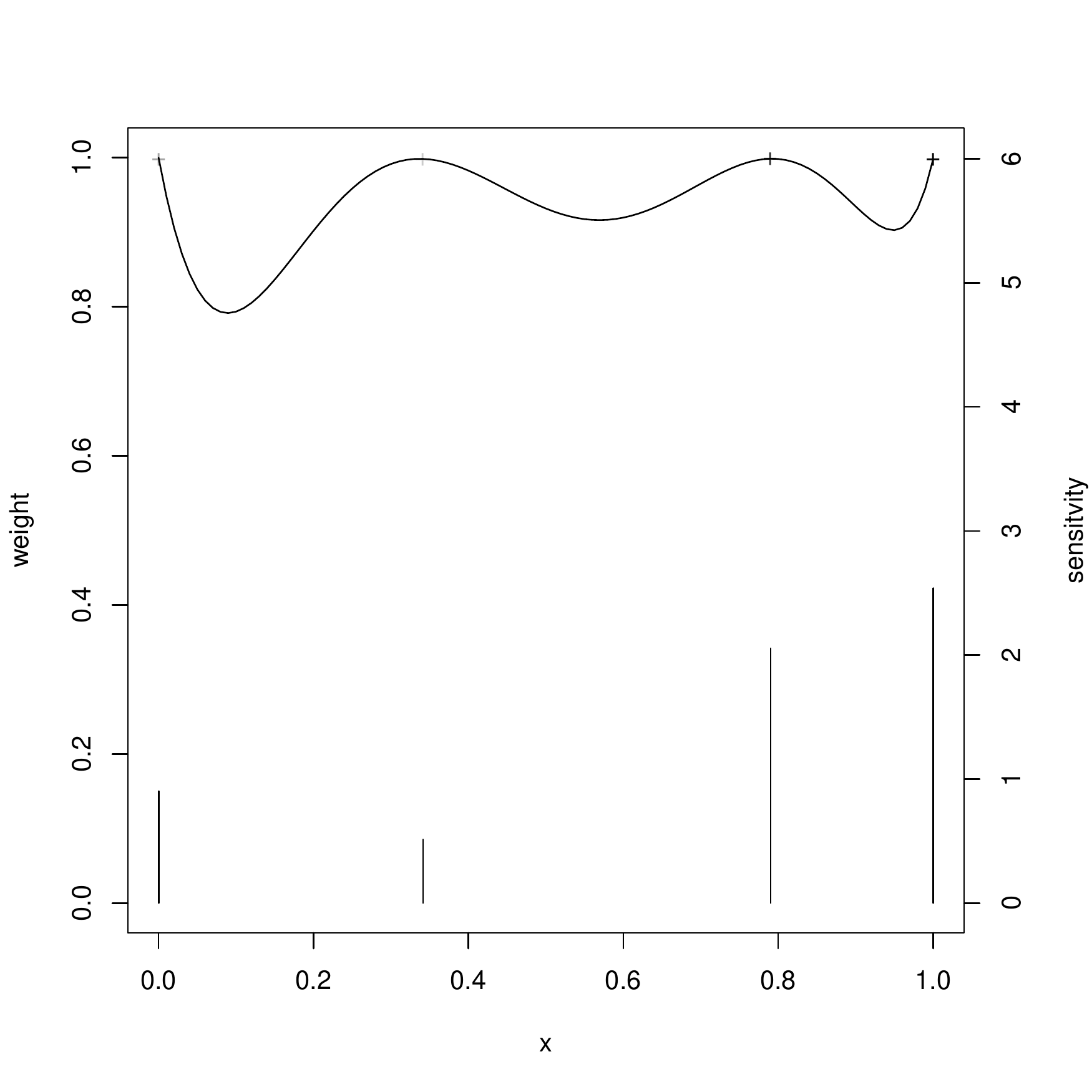}
\end{minipage}
\caption{Design points (first column), weights (second column), sensitivity function (continuous line) and weights (bars) of the $D_s$-optimal design for $\beta_1,\dots,\beta_6$.}\label{fig:Fed}
\end{figure}
This example has already been generalized in \cite{Perrone_16} where various dependences through copula functions have been introduced and the corresponding $D$-optimal designs have been computed. 
In order to illustrate the usage of $D_s$-optimality in this context, let us assume the dependence between $Y_1$ and $Y_2$ described by a Clayton copula with $\alpha_1 = 18$, corresponding to a Kendall's $\tau$ (see equation (\ref{eq:tau})) value of $0.9$.

Even though the low losses in $D$-efficiency reported in \cite{Perrone_16} suggest that the impact of the assumed dependence is completely negligible, one might aim at verifying whether the information related to the dependence structure is only carried by the estimation of $\alpha_1$.
Essentially, one might focus on the six marginal parameters entirely disregarding the estimation of the dependence parameter $\alpha_1$.
This can be done in practice by applying the $D_s$-optimality to the parameter vector $\boldsymbol{\beta}$. 

Figure \ref{fig:Fed} shows the $D_s$-optimal design corresponding to this case.
Comparing the $D$-optimal design of the product copula, assuming no dependence, with the $D_s$-optimal design for only the vector $\boldsymbol{\beta}$, the loss in $D_s$-efficiency is of $8\%$. This shows that the dependence structure itself can substantially affect the design even if the dependence parameter $\alpha_1$ is ignored in the estimation.

In more complex models, a similar approach can be used to identify informative designs to specific properties of interest.
In the following, we highlight the usefulness of flexible copula models through the application of the $D_s$-criterion to a subclass of meaningful model parameters. We construct in this way designs which better reflect the strength and the structure of a specific dependence and can then be used to discriminate between classes of copulas.

\section{Bivariate binary case}
\subsection{Copulas: combinations and measure of association}
As already mentioned above, the problem of specifying a probability model for dependent random variables can be simplified by expressing the corresponding 2-dimensional joint distribution $F_{Y_1Y_2}$ in terms of its two margins $F_{Y_1}$ and $F_{Y_2}$, 
and an associated {2-copula} (or dependence function) $C$, implicitly defined through the functional identity stated by Sklar's Theorem \cite{sklar_59}.
Copula theory allows the practitioner to gain in flexibility as for example any finite \textit{convex linear combination} of 2- copulas $C_i$'s is itself a 2-copula.
In particular, for $k \in \mathbb{N}$, let $C$ be given by
\begin{equation}
C(u,v)= \sum_{i=1}^{k} \lambda_i C_i (u,v),
\end{equation}
where $\lambda_i \geq 0$ for all indexes, and 
$\sum_{i=1}^{k} \lambda_i = 1$. Then, $C$ is a 2-copula.

Another useful property of copulas is that considering $Y_1$ and $Y_2$ two continuous random variables whose copula is $C_{\alpha_1}$, the measure of association Kendall's $\tau$ is related to the expectation of the random variable $W = C_{\alpha_1}(U,V)$, which is
\begin{equation}
\label{eq:tau}
\tau = 4 \int\limits_{I}\int\limits_{I} C_{\alpha_1}(u,v) d C_{\alpha_1}(u,v) - 1.
\end{equation}
The relation in Equation (\ref{eq:tau}) results in a correspondence between the copula parameter $\alpha_1$ and a fixed $\tau$ value.
Such a relationship can be used in the construction of extremely flexible models, as shown in the next example.

\subsection{The example}
We analyze an example with potential applications in clinical trials already examined in \cite{denman+al_11} and \cite{Perrone_16}. 
We consider a bivariate binary response $(Y_{i1}, Y_{i2})$, $i=1, \ldots, n$ with four possible outcomes $\{ (0,0),(0,1),(1,0),(1,1)\}$ where $1$ usually represents a success and $0$ a failure (of, e.g., a drug treatment where $Y_1$ and $Y_2$ might be efficacy and toxicity).
For a single observation denote the joint probabilities of $Y_1$ and $Y_2$ by $p_{y_1,y_2} = \mathbb{P} (Y_1 = y_1, Y_2 = y_2)$ for $(y_1,y_2 \in\{0,1\})$.
Now, define
\begin{equation}
\begin{array}{cccccc}
  p_{11} & = & C_{\boldsymbol{\alpha}} (\pi_1, \pi_2), & p_{10} & = & \pi_1 - p_{11}, \\
  p_{01} & = & \pi_2 - p_{11}, &  p_{00} & = & 1 - \pi_1 - \pi_2 + p_{11}. 
\end{array}
\end{equation}

A particular case of the introduced model has already been analyzed in \cite{heise+m_1996}.
In that work, the authors assume the marginal probabilities of success given by the models
\begin{equation}
\log \left( \dfrac{\pi_i}{1 - \pi_i} \right) = \beta_{i1} +  \beta_{i2} x, \qquad i=1,2
\end{equation}
with $x \in [0,10]$ and `localized' parameters $\tilde{\boldsymbol{\beta_1}}=(-1, 1)$ and $\tilde{\boldsymbol{\beta_2}}=(-2, 0.5)$. 
Let us now allow the strength of the dependence itself be dependent upon the regressor $x$.
As in our context only positive associations make sense we consider in the following the corresponding Kendall's $\tau$ modeled by a logistic:
\[\tau(x, \alpha_1) = \dfrac{e^{\alpha_1 x - c}}{1 + e^ {\alpha_1 x - c}}, 
\]
where $c$ is a constant chosen such that $\tau$ takes values in $[\epsilon,1]$ for $\alpha_1 \in [0,1]$. For our computations we choose $\epsilon=0.05$ and we select three values for $\alpha_1$ such that the $\tau$ ranges are $I_1 = [0.05,0.3], \; I_2= [0.05,0.9]$, and $I_3= [0.05,0.95]$.

Then, using the relationship from equation (\ref{eq:tau}) that associates the Kendall's $\tau$ with the copula parameter, we model $p_{11}$ by pair convex combinations of Joe, Frank, Clayton, and Gumbel copulas by linking the two copulas $C_1$ and $C_2$ at the same $\tau$ values through the functions $h_1$ and $h_2$:
  \[C(\pi_1,\pi_2; \alpha_1, \alpha_2) = \alpha_2 C_1(\pi_1,\pi_2; \; h_1(x,\alpha_1)) + (1-\alpha_2) C_2(\pi_1,\pi_2; \, h_2(x,\alpha_1)).\]
  Notice that the construction is more general and any convex combination of standard copulas from the R package 'copula' can be considered through the package 'docopulae' (\cite{Docopulae}).

\begin{figure}
\begin{center}
\begin{tabular}{c c}
\includegraphics[scale=0.38]{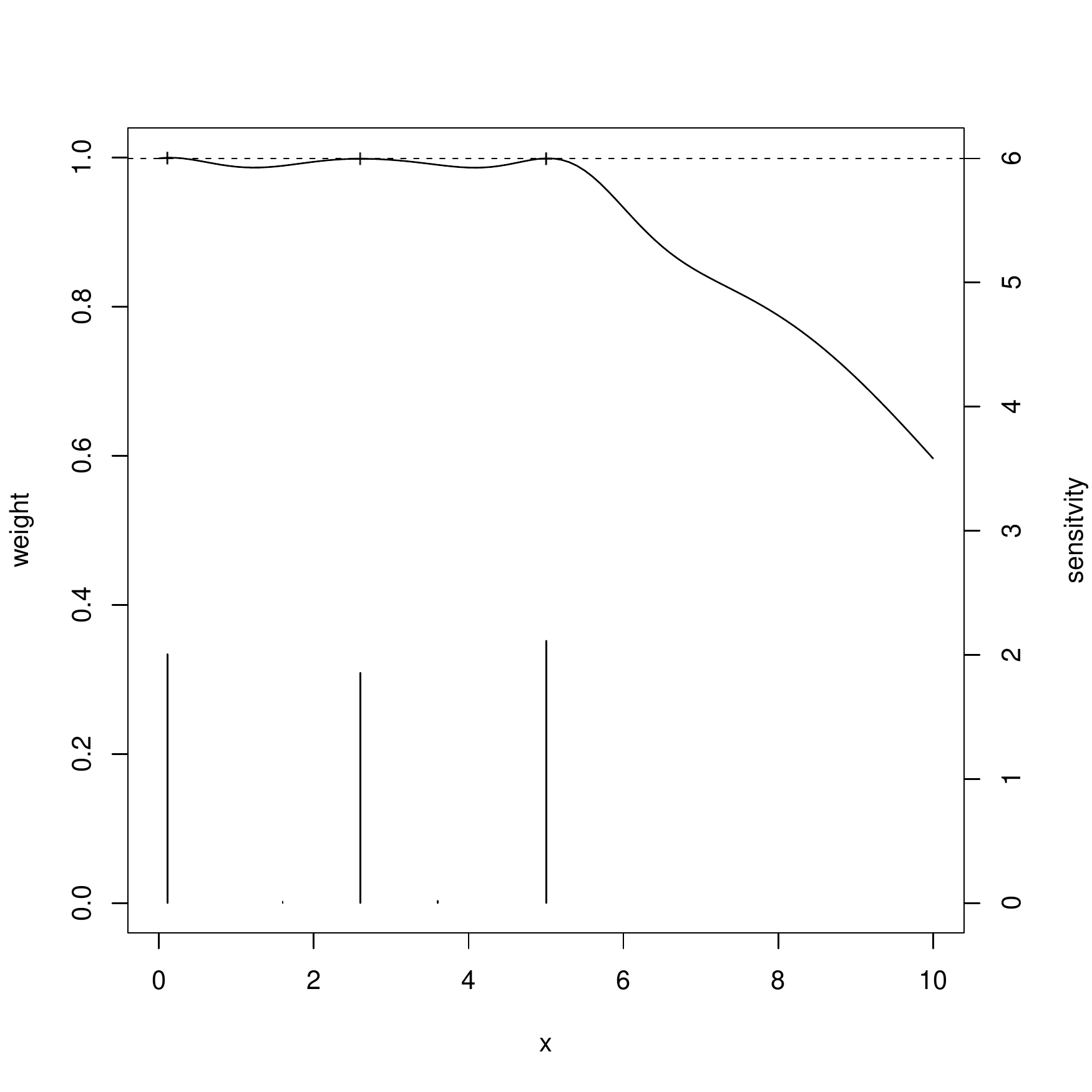}
\quad & \quad
\includegraphics[scale=0.38]{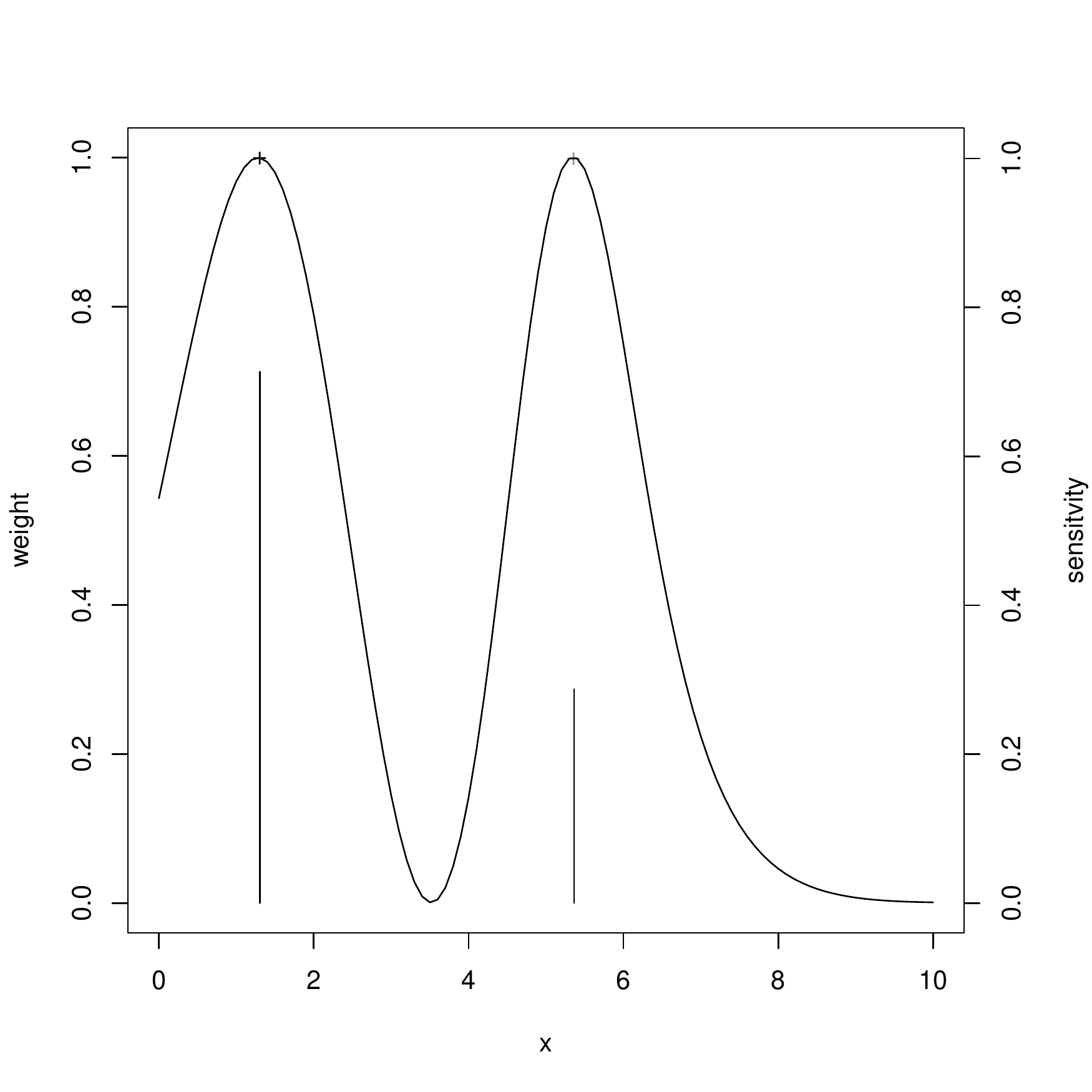} \\
\includegraphics[scale=0.38]{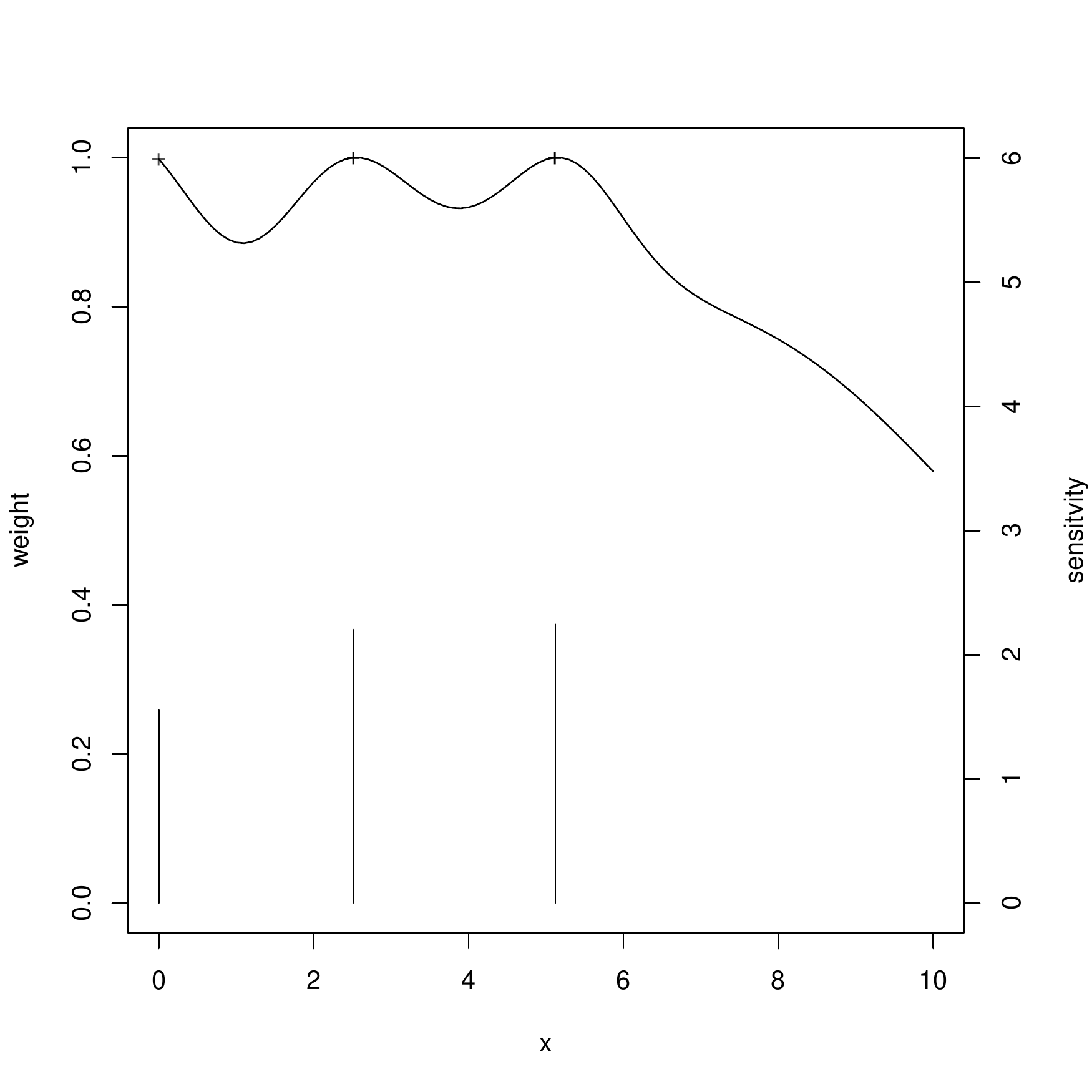}
\quad & \quad
\includegraphics[scale=0.38]{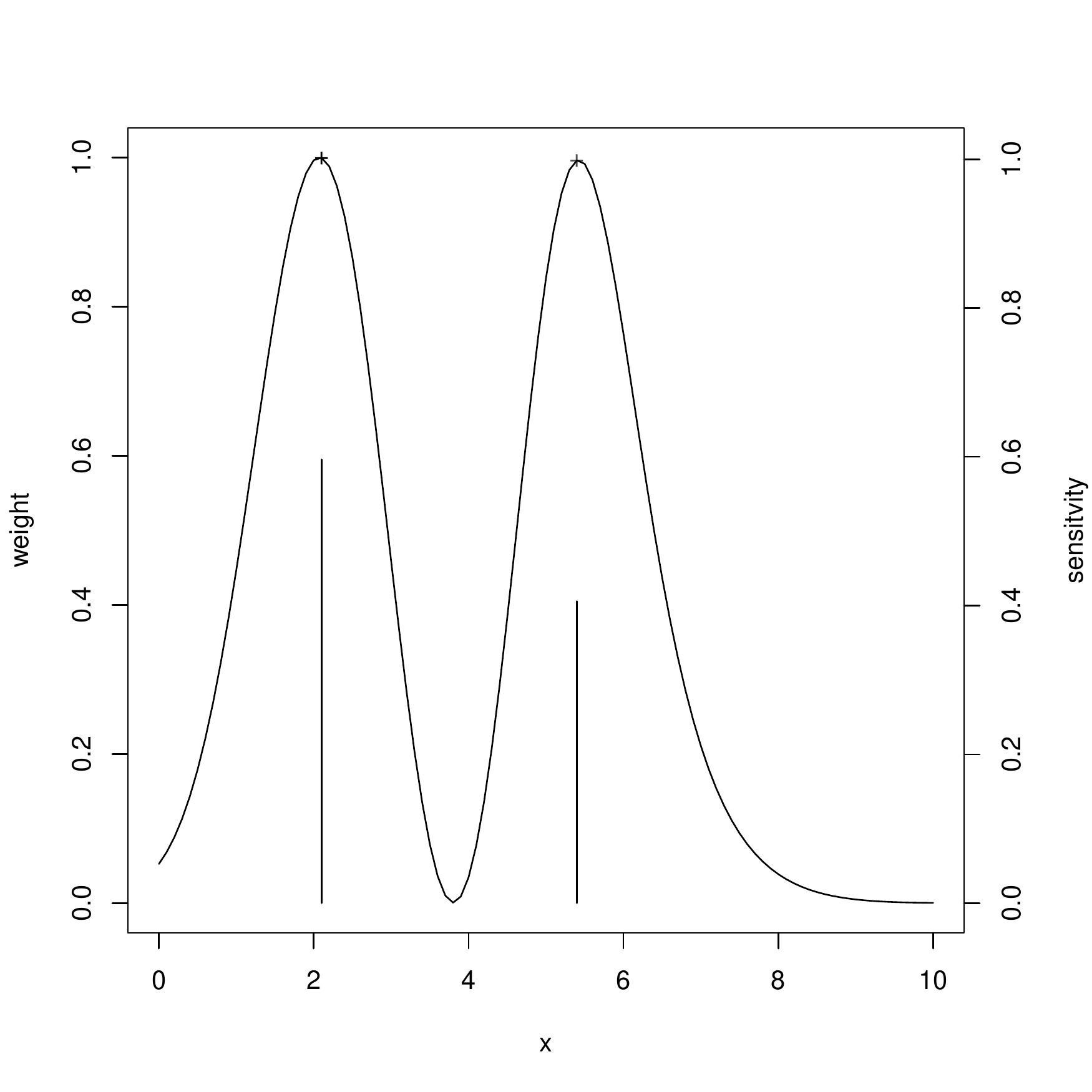}
\end{tabular}
\caption{Sensitivity functions (continuous lines) and weights (bars) for $D$-optimal (left column) and $D_s$-optimal  (right column) designs for Clayton-Gumbel (first line) and Frank-Gumbel (second line) with $\tau \in I_2=[0.05,0.9]$ and $\alpha_2=0.5$.}\label{fig:Biv_case}
\end{center}
\end{figure}

\begin{table} \centering
\medskip
\begin{tabular}{|c|c|c|c||c|c|c|}
\hline
  \multicolumn{4}{|c||}{{Joe - Frank}} &  \multicolumn{3}{|c|}{{Clayton - Gumbel}}\\
  \hline
 $\tilde{\alpha}_2$ & $\tau \in I_1$ & $\tau \in I_2$ & $\tau \in I_3$ & $\tau \in I_1$ & $\tau \in I_2$ & $\tau \in I_3$\\ 
 \hline
0.1 & 34.94 & 38.80 & 41.37 & 49.85 & 49.45 & 45.10 \\ 
 \hline
 0.5 & 42.36 & 38.20 & 41.83 & 43.65 & 39.27 & 39.03 \\ 
   \hline
 0.9 & 55.11 & 47.23 & 44.15 & 37.87 & 34.65 & 37.78 \\ 
   \hline\hline
     \multicolumn{4}{|c||}{{Joe - Clayton}} &  \multicolumn{3}{|c|}{{Frank - Gumbel}}\\
  \hline
 $\tilde{\alpha}_2$ & $\tau \in I_1$ & $\tau \in I_2$ & $\tau \in I_3$ & $\tau \in I_1$ & $\tau \in I_2$ & $\tau \in I_3$\\ 
 \hline
0.1  & 35.92 & 36.35 & 39.01 & 47.13 & 48.29 & 46.17 \\ 
 \hline
 0.5  & 45.37 & 43.17 & 45.53 & 37.65 &  34.41 & 34.37\\ 
   \hline
 0.9 & 49.92 & 48.72 & 45.36 & 38.51 & 34.19 & 36.26 \\ 
   \hline\hline
\end{tabular}
\caption{\label{tab1} Losses in $D_s$-efficiency in percent for $I_1 = [0.05,0.3], \; I_2= [0.05,0.9]$, and $I_3= [0.05,0.95]$.}
\end{table}
In this model, the impact of the dependence structure and the association level is reflected by two different parameters, as the $\alpha_1$ parameter is only related to the measure of 
association Kendall's $\tau$, while the $\alpha_2$ parameter is strictly related to the 
structure of the dependence. Therefore, applying the $D_s$-criterion on $\alpha_2$, we find a design for discriminating, in this specific model, between the two copulas considered. 

We compare the design obtained for different $\tau$ intervals and localized values for $\alpha_2$ with the $D$-optimal design obtained for the same localized values (Figure \ref{fig:Biv_case}). 
Analyzing the rather high losses in $D_s$-efficiency reported in Table \ref{tab1}, it shows that the $D$-criterion alone is not sufficient when we require information about the structure of the model.

In this scenario, an interesting question is whether the obtained $D_s$-optimal designs are robust with respect to the initial model assumptions. 
To analyze this aspect, we computed the $D_s$-efficiencies for cross-comparisons of $D_s$-optimal designs. 
In Table \ref{tab2}, the results for $\tau \in I_2$ and $\tilde{\alpha}_2 = 0.5$ are reported (see Figure \ref{fig:Biv_case}, also). Looking at the table, one can notice that the losses correspondent to the assumed combination Clayton-Gumbel are in general lower, not exceeding $16\%$. 
This means that such a combination provides good results in order to discriminate between all the considered dependences. Further studies in this direction would lead to the development of new design techniques to construct robust and stable designs for discrimination between various classes of dependences.

\begin{table}[!ht]
\begin{center}{
\begin{tabular}{l||c|c|c|c|}
\cline{2-5}
& \multicolumn{4}{|c|}{Assumed Copula } \\
\hline
\multicolumn{1}{|c||}{True Copula} &  \multicolumn{1}{|c|}{{C-G}} & \multicolumn{1}{|c|}{{F-G}} & \multicolumn{1}{|c|}{{J-C}} &  \multicolumn{1}{|c|}{{J-F}}\\
\hline
\multicolumn{1}{|c||}{Clayton - Gumbel (C-G)} & ${0.00}$  & ${28.44}$ & ${7.43}$ & ${19.07}$ \\
\hline
\multicolumn{1}{|c||}{Frank - Gumbel (F-G)} & ${16.09}$  & ${0.00}$ &  ${30.17}$ & ${19.51}$ \\
\hline
\multicolumn{1}{|c||}{Joe - Clayton (J-C)} & ${4.25}$  & ${34.27}$ & ${0.00}$ & ${13.51}$ \\
\hline
\multicolumn{1}{|c||}{Joe - Frank (J-F)} & ${15.13}$  & ${13.97}$ & ${9.52}$ & ${0.00}$ \\
\hline
\hline
\end{tabular}
}
\end{center}
\caption{Losses in $D_s$-efficiency in percent for $\tau \in I_2$ and $\tilde{\alpha}_2=0.5$ by comparing the true copula model with the assumed one.}
\label{tab2}
\end{table}

\section{Bivariate Weibull function}
In this section we extend an example originally reported in \cite{Kim_15}. %
After providing a brief overview of the theoretical framework, we construct original asymmetric copula models and we apply $D_s$-optimality to discriminate between symmetric and asymmetric scenarios. 
\subsection{Copulas and exchangeability}
A copula $C$ is said to be \emph{exchangeable} (or \emph{symmetric}) if it does not change under any permutation of its arguments.
In particular, if $(U,V)$ is a random pair distributed according to an exchangeable copula $C$, then
$$
\mathbb{P}(V\le v\mid U\le u)=\mathbb{P}(U\le v\mid V\le u),
$$
for all $(u,v)\in[0,1]^2$. Consequently, the conditional distributions of $(V\mid U\le u)$ and $(U\mid V\le u)$ are equal. This indicates that a causality relationship between $U$ and $V$ leads to non-exchangeability. 

Possible ways of quantifying non-exchangeability in copula models have been provided in the literature (\cite{klement+m_06,Nelsen_07}).
Although some classes of bivariate copulas can directly deal with non-exchangeability (\cite{Capera_00,Charpentier_14,Klement_05,DeBaets_07}), many other copulas largely used in modeling belong to the class of exchangeable ones. 
An example is given by the well-known family of Archimedean copulas (see \cite{DurSem15,Genest_86,nelsen_06}), which are not suitable to model many situations that might arise in real scenarios (see, for instance, \cite{Genest_13}).

To overcome these restrictions, a possibility is to apply transformations which commute exchangeable copulas into non-exchangeable ones (\cite{Durante_2007,Frees_98,Khoudraji_95}).
In the next example, we apply the Khoudraji's asymmetrization described in \cite{Khoudraji_95}.
In particular, we modify a given exchangeable copula $C_{\alpha_1}$, with parameter $\alpha_1$, into the copula $C=C_{\alpha_1,\alpha_2,\alpha_3}$ defined, for every $(u,v)\in [0,1]^2$, by
\begin{equation}
\label{Eq:AC}
C(u,v) = u^{\alpha_2} v^{\alpha_3} C_{\alpha_1}(u^{1-\alpha_2}, v^{1-\alpha_3}),
\end{equation}
where $\alpha_2,\;\alpha_3 \in[0,1]$. For $\alpha_2\neq \alpha_3$, $C$ is non-exchangeable. 

First investigations on the changes in the geometry of the $D$-optimal design for such transformations have been carried out in \cite{Durante_16}, where a theoretical overview of exchangeability in the copula theory is also given. 
In the following we instead present the usage of the $D_s$-optimality to discriminate between symmetric and asymmetric models.

\subsection{The Weibull case}
We now analyze an example originally reported in \cite{Kim_15}. We assume two dependent binary outcomes, $U$ and $V$, for two system components, respectively. Considering $0$ indicating no failure and $1$ indicating failure, the outcome probabilities given a stress $x$ can be written as: 
$$p_{uv}(x, \boldsymbol{\gamma}) = \mathbb{P}(U=u,V=v\mid x, \boldsymbol{\gamma}),$$
with  $u,v \in \{ 0,1\}$ and where $\boldsymbol{\gamma}$ denotes a vector of all the model parameters. 

Let $Y$ and $Z$ denote the amount of damage on component 1 and component 2, respectively, and let $f(y,z\mid x,\boldsymbol{\gamma})$ be the bivariate Weibull regression model. Suppose that failures are defined by dichotomizing damage measurements $Y$ and $Z$: 
\begin{equation}
\begin{array}{c}
U = \left\{ \begin{array}{cc} 0 & \text{ (no failure for component 1), if } Y < \zeta_1, \\ 1 & \text{ (failure for component 1), otherwise}\end{array}\right. \\
 \\
V = \left\{ \begin{array}{cc} 0 & \text{ (no failure for component 2), if } Z < \zeta_2, \\ 1 & \text{ (failure for component 2), otherwise}\end{array}\right.
\end{array}
\end{equation}
 where $\zeta_1$ and $\zeta_2$ are predetermined cut-off values. Then, the probabilities of success and failure are:
 \begin{equation}
\begin{array}{c}
p_{00} = \int_0^{\zeta_1} \int_0^{\zeta_2} f(y,z\mid x,\boldsymbol{\gamma}) \,d y\,d z, \quad p_{01} = \int_0^{\zeta_1} \int_{\zeta_2}^{\infty} f(y,z\mid x,\boldsymbol{\gamma}) \,d y\,d z,  \\
\\
p_{10} = \int_{\zeta_1}^{\infty} \int_0^{\zeta_2} f(y,z\mid x,\boldsymbol{\gamma}) \,d y\,d z, \quad 
p_{11} = \int_{\zeta_1}^{\infty} \int_{\zeta_2}^{\infty} f(y,z\mid x,\boldsymbol{\gamma}) \,d y\,d z.
\end{array}
 \end{equation}
 Now, considering $f(y,z\mid x,\boldsymbol{\gamma})$ defined as follows:
{\scriptsize
\[
f(y,z) = \left\{ \begin{array}{lc} 
 \beta_1(\beta_3 + \beta_5) \kappa^2 (yz)^{\kappa - 1} \text{exp}\{ -(\beta_3 + \beta_5) z^{\kappa} - (\beta_1 + \beta_2 - \beta_5) y^{\kappa}\} & \text{for } 0 < y < z < \infty; \\
\beta_2(\beta_3 + \beta_4) \kappa^2 (yz)^{\kappa - 1} \text{exp}\{ -(\beta_3 + \beta_4) y^{\kappa} - (\beta_1 + \beta_2 - \beta_4) z^{\kappa}\} & \text{for } 0 < z < y < \infty; \\
\beta_3 \kappa (y)^{\kappa - 1} \text{exp}\{ -(\beta_1 + \beta_2 + \beta_3) & \text{for } 0 < y = z < \infty.
\end{array}\right. 
 \]
}
The marginal survival functions of the bivariate Weibull density are weighted univariate Weibull survival functions:
{\footnotesize
\[ \mathbb{P}(Y \geq y) = \dfrac{\beta_2} {\beta_1 + \beta_2 - \beta_4} \text{exp} \{ - (\beta_3 + \beta_4)  y^{\kappa}\} + \left(1- \dfrac{\beta_2}{\beta_1 + \beta_2 - \beta_4}\right) \text{exp} \{ - (\beta_1 + \beta_2 + \beta_3) y^{\kappa} \}\]
\[ \mathbb{P}(Z \geq z) = \dfrac{\beta_1} {\beta_1 + \beta_2 - \beta_5} \text{exp} \{ - (\beta_3 + \beta_5)  z^{\kappa}\} + \left(1- \dfrac{\beta_1} {\beta_1 + \beta_2 - \beta_5}\right) \text{exp} \{ - (\beta_1 + \beta_2 + \beta_3) z^{\kappa} \}\]
}
In \cite{Kim_15}, the authors set $\zeta_1=0.8$ and $\zeta_2 = 0.7$. Moreover, they consider the following predictor functions:
\begin{equation}
\left\{ \begin{array}{rll} 
        -\log (\beta_3 + \beta_5) & = & \theta_0 + \theta_1 x , \\
        -\log (\beta_3 + \beta_4) & = & \theta_0 + \theta_2 x , \\
        -\log (\beta_1 + \beta_2 + \beta_3) & = & \theta_0 + \theta_3 x.
 \end{array}\right.
\end{equation}
with $x\in[0,1]$.
\begin{figure}
\begin{center}
\begin{tabular}{c c}
\includegraphics[scale=0.38]{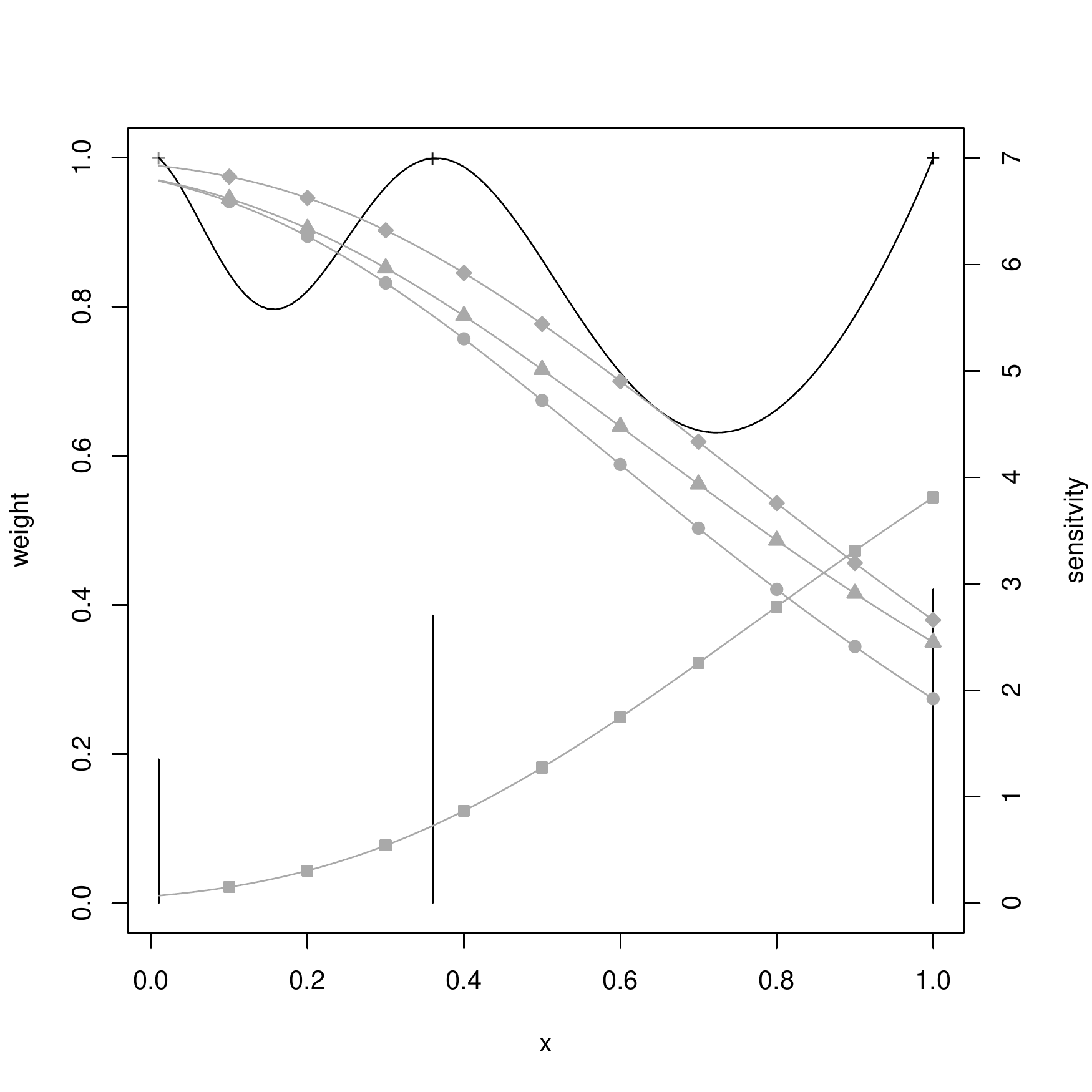}
\quad & \quad
\includegraphics[scale=0.38]{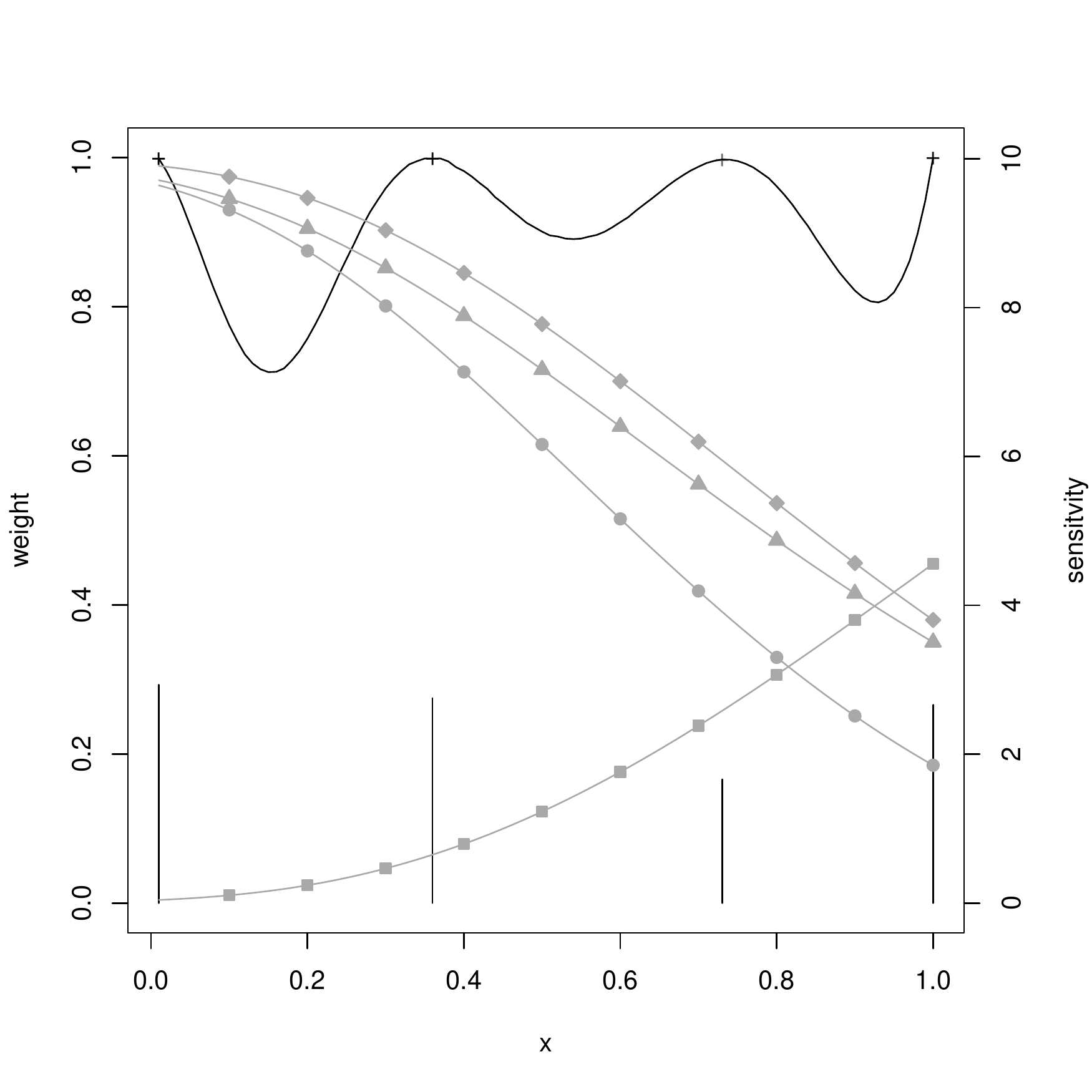} \\
\end{tabular}
\caption{Sensitivity functions (continuous lines) and design weights (bars) of the $D$-optimal design for the Weibull case as reported in \cite{Kim_15} (left), and for asymmetric Clayton with $(\tilde{\alpha}_1,\tilde{\alpha}_2,\tilde{\alpha}_3)= (1.5,0.4,0)$ (right); \textcolor{silver}{$\CIRCLE$}$,p_{00}$; \textcolor{silver}{$\blacksquare$}$,p_{11}$; \textcolor{silver}{$\blacklozenge$}$,p_{0.}$; \textcolor{silver}{$\blacktriangle$}$,p_{.0}$ }
\label{fig:Wei_case}
\end{center}
\end{figure}
\begin{figure}
\begin{center}
\begin{tabular}{c c}
\includegraphics[scale=0.38]{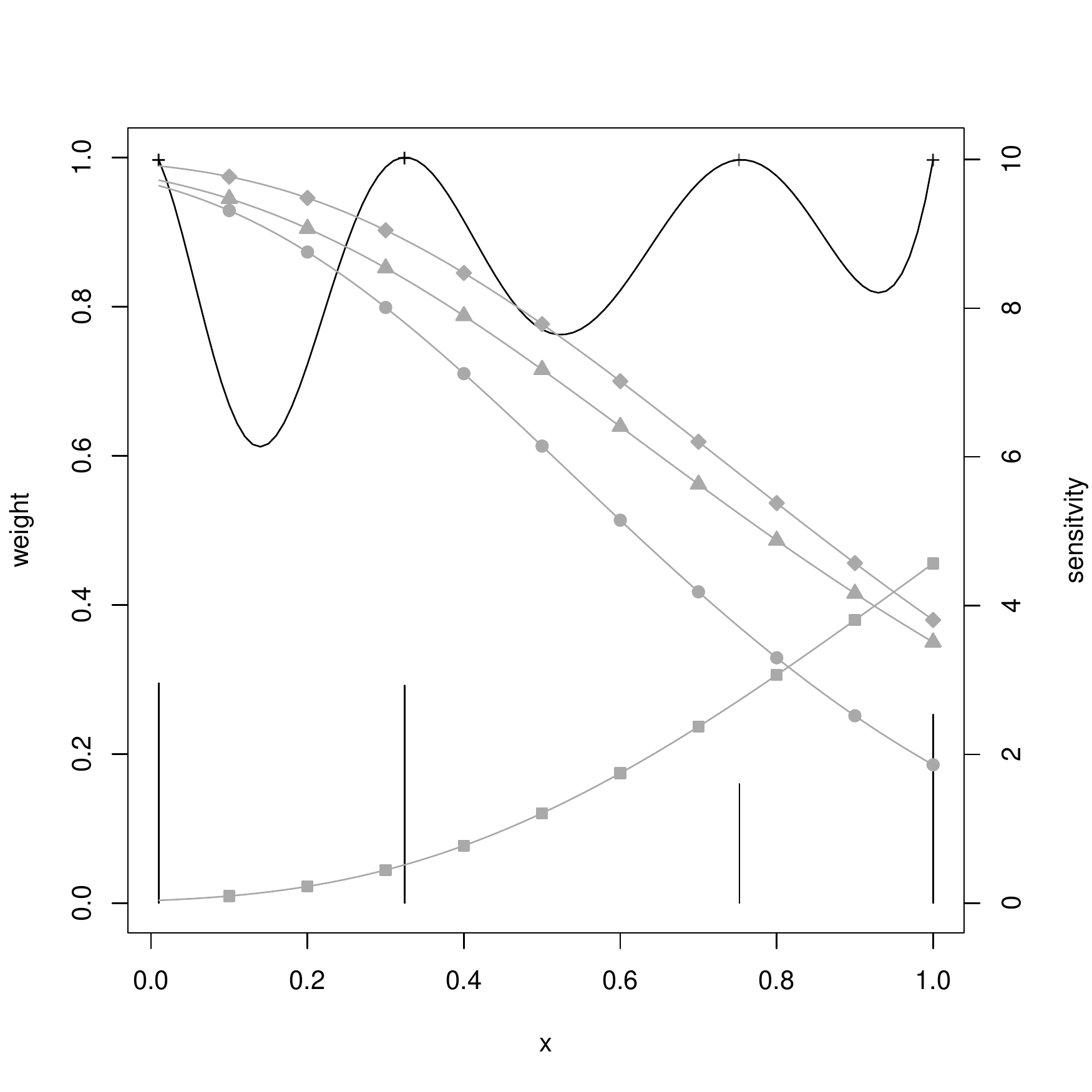}
\quad & \quad
\includegraphics[scale=0.38]{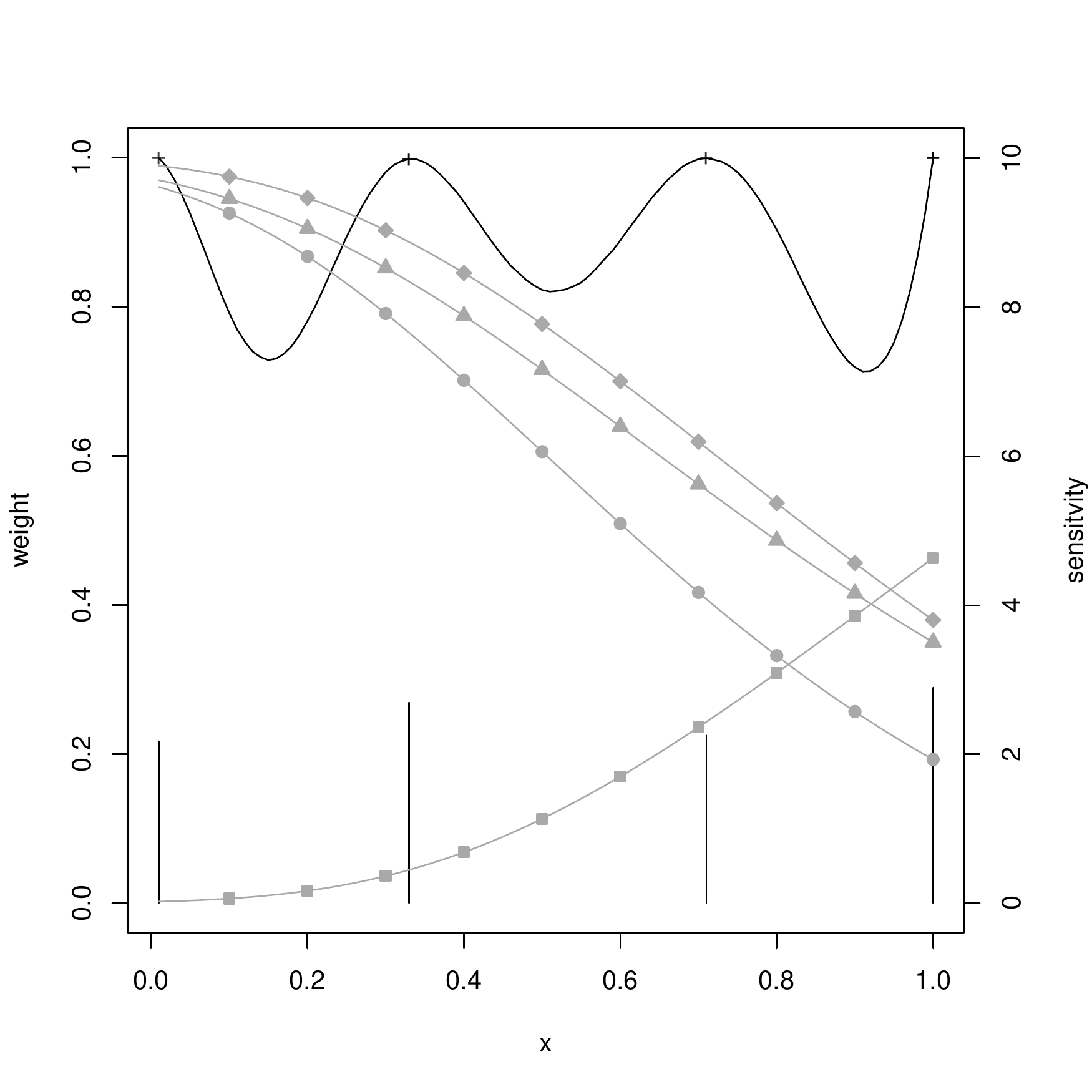}\\
\includegraphics[scale=0.38]{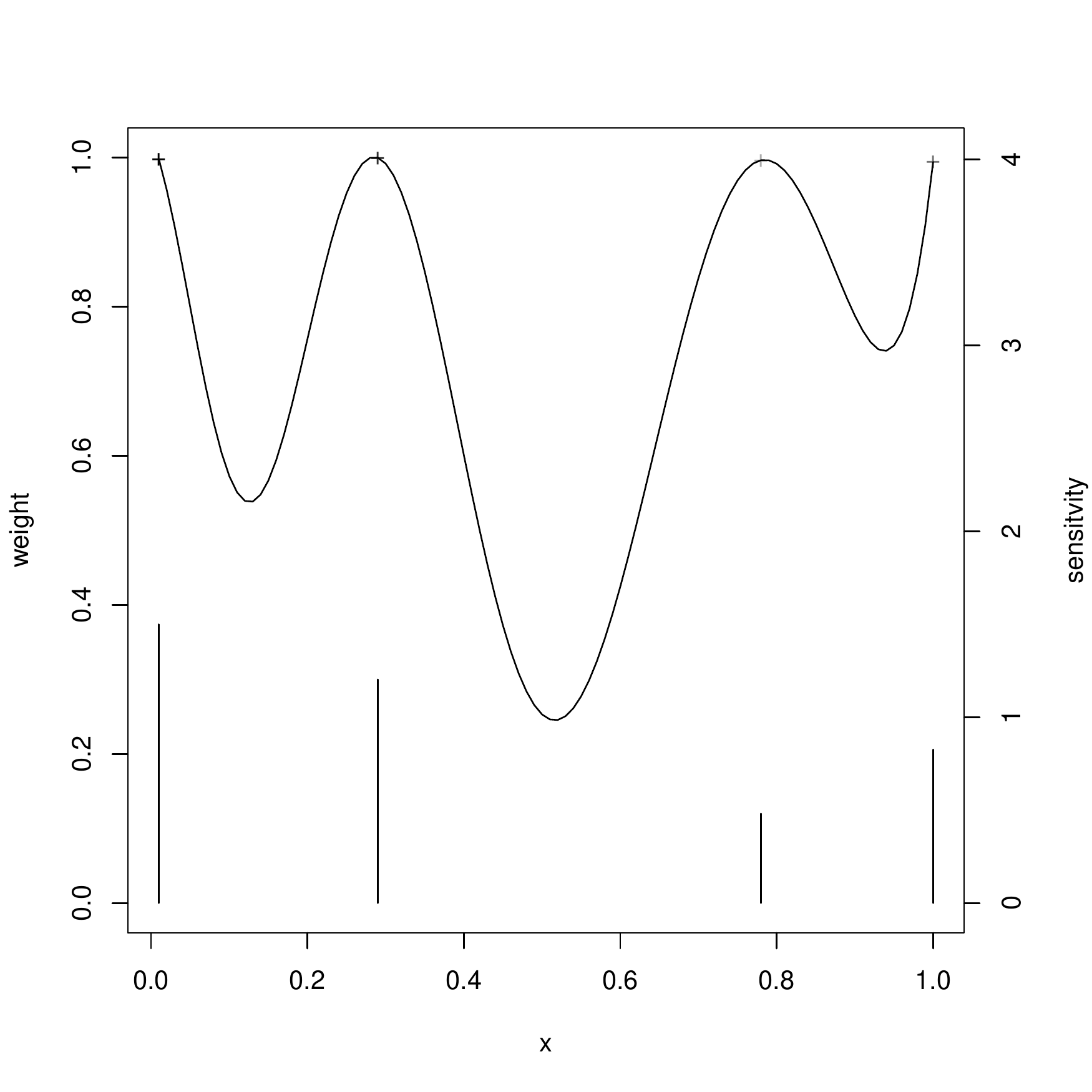}
\quad & \quad
\includegraphics[scale=0.38]{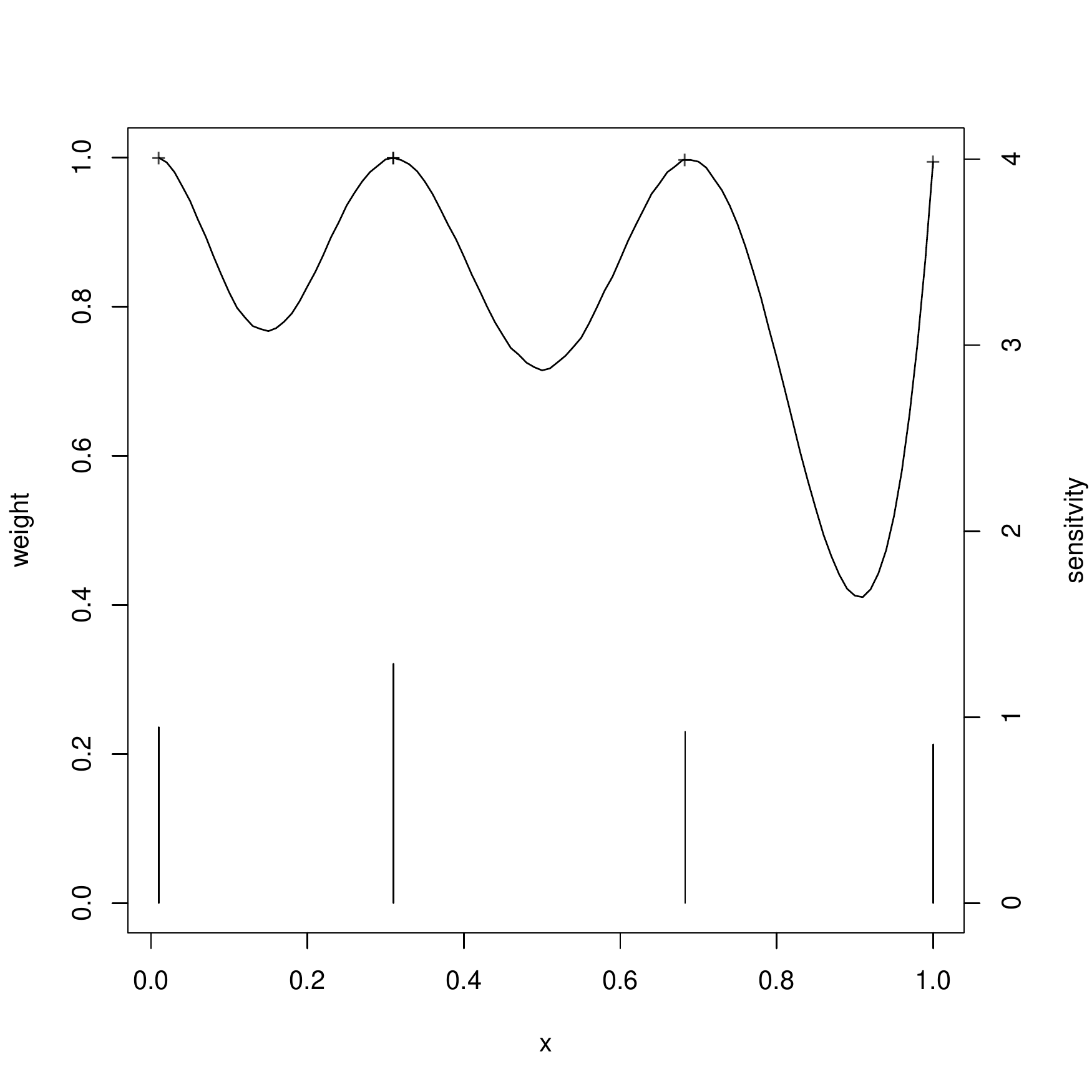}
\end{tabular}
\caption{Sensitivity functions (continuous lines) and design weights (bars) of $D$-optimal designs (first row) and $D_s$-optimal designs (second row) for the Weibull case for asymmetric Clayton with $(\tilde{\alpha}_1,\tilde{\alpha}_2,\tilde{\alpha}_3)= (2,0.4,0.2)$ (left column), and for $(\tilde{\alpha}_1,\tilde{\alpha}_2,\tilde{\alpha}_3)= (3.6,0.6,0)$ (right column);  \textcolor{silver}{$\CIRCLE$}$,p_{00}$; \textcolor{silver}{$\blacksquare$}$,p_{11}$; \textcolor{silver}{$\blacklozenge$}$,p_{0.}$; \textcolor{silver}{$\blacktriangle$}$,p_{.0}$}\label{fig:DsWei_case}
\end{center}
\end{figure}
In \cite{Kim_15}, the asymmetry in the causality has been reflected by different cut points, e.g., unequal values for $\zeta_1$ and $\zeta_2$, and different initial failure rates $\beta_1$ and $\beta_2$ as well as different coefficients $\theta_1$ and $\theta_2$ of the predictor. 

In our example, we additionally allow asymmetry of the phenomenon to appear in the dependence structure.
In particular, such an asymmetry is introduced through the transformation presented in equation (\ref{Eq:AC}), adding new parameters in the process. 

Going into details, we introduce two parameters $\nu_1$, and $\nu_2$ such that the following is satisfied:
 \[ \left\{ \begin{array}{l} 
 \theta_1 = \theta_2 + \nu_1,\\ \beta_1 = \beta_2 + \nu_2. \end{array}\right.\]
 The vector $(\nu_1, \nu_2)$ then quantifies the dissimilarity of the margins. For our study, we assume the joint dependence to be described by the asymmetric Clayton copula with three parameters $\alpha_1, \; \alpha_2$ and $\alpha_3$, constructed according to equation (\ref{Eq:AC}).
In this context, we apply $D_s$-optimality to the parameters $\boldsymbol{\mu} = (\nu_1, \nu_2, \alpha_2, \alpha_3)$ which denote the total asymmetry of the phenomenon, both from the marginals and the joint dependence. In such a way, we find designs which are more informative to the asymmetry and are then suitable to discriminate between exchangeable models and non-exchangeable ones. 
The used parameter setting corresponds to two Kendall's tau values: 0.5 and 0.25, respectively. The initial values of the parameters ${\alpha}_1, {\alpha}_2,$ and ${\alpha}_3$ are the same as used in \cite{Durante_16}, while the other parameter values are $\tilde{\theta}_0=-2,\; \tilde{\theta}_2=5,\; \tilde{\theta}_3=2,\; \tilde{\nu}_1=-1,\; \tilde{\nu}_2=0.1,\; \tilde{\beta}_2=0.2,$ and $\tilde{\kappa}=2$.

The $D$-optimal designs obtained spread weight to four design points, slightly differing in their distribution.
Figure \ref{fig:Wei_case} shows a representative design for our model side by side with the $D$-optimal design for the Weibull case as reported in \cite{Kim_15}.
The maximal and minimal values of the loss in $D$-efficiency by comparing the design reported in \cite{Kim_15} and the $D$-optimal designs for our models are reported in Table \ref{tab4}. A full table with the losses of such comparison for each set of initial values of ${\alpha}_1, \; {\alpha}_2,$ and $ {\alpha}_3$ is available in the supplementary material. The results suggest that in every case it would be advantageous to choose one of our models as generally more informative and robust.

\begin{table}[b]
\centering
\begin{tabular}{|c||r|r|r|r|}
\cline{2-5}
  \multicolumn{1}{c||}{} & \multicolumn{4}{|c|}{True Model} \\
  \cline{2-5}
  \multicolumn{1}{c||}{} & \multicolumn{2}{|c|}{ Weibull} & \multicolumn{2}{c|}{Our Models}\\
  \hline
Assumed Model & min & max & min & max \\
\hline
Weibull & 0.00 & 0.00 & 17.78 & 71.65 \\
\hline
Our Models & 9.43 & 10.18 & 0.00 & 3.37 \\
   \hline
   \hline
\end{tabular}
\caption{\label{tab4} Losses in $D$-efficiency in percent for crossed comparison between the optimal design find for the Weibull model as reported in \cite{Kim_15} and all our models.}
\end{table}

We are now interested in verifying whether the $D$-optimal design is informative enough to discriminate between asymmetry and symmetry. To this aim, we compare $D_s$-optimal designs for $\boldsymbol{\mu}$ to the corresponding $D$-optimal designs (Figure \ref{fig:DsWei_case}). In this case, the loss in $D_s$-efficiency never exceeds $5\%$. In contrast to the binary case, such a result indicates that the $D$-optimal design is already quite adequate for discriminating between symmetric and asymmetric models.

\section{Discussion}
In this paper we embed the issue of the choice of the copula in the framework of discrimination design. We present a new methodology based on the $D_s$-optimality to construct design that discriminate between various dependences. 
Through some examples we highlight the strength of the proposed technique due to the usage of the copula properties. 
In particular, the proposed approach allows to check the robustness of the $D$-optimal design in the sense of discrimination and to construct more informative designs able to distinguish between classes of dependences.

All the shown results are obtained by the usage of the R package 'docopulae' (\cite{Docopulae}). Although we here compare just few possible dependences, the general construction is much wider. The R package 'docopulae' allows the interested reader to run designs assuming a broad variety of dependence structures. It then provides a strong computational tool to the usage of copula models in real applications.

The innovative approach we present in this paper is promising as it breaks new ground in the field of experimental design.
In the future, we aim at generalizing other discrimination criteria such as $T-$optimality and $KL-$optimality to flexible copula models (\cite{dette+t_09,ucinski_05,lopez-fidalgo_2007}). Furthermore, powerful compound criteria might be developed for such models (see, for instance, \cite{atkinson_08,dette_93,tommasi_09}).
In addition, the construction of multistage design procedures that allow for discrimination and estimation might be of great interest in special applications such as clinical trial studies (\cite{dragalin_08,mueller_96}).

\section*{Acknowledgments}
This work has been supported by the project ANR-2011-IS01-001-01 ``DESIRE" and Austrian Science Fund (FWF) I 833-N18.

We thank F. Durante, L. Pronzato, J. Rendas and E.P. Klement for fruitful discussions.


\end{document}